# Detection of bromochloro alkanes in indoor dust using a novel CP-Seeker data integration tool

Thomas J. McGrath[a,b], Julien Saint-Vanne[a], Sébastien Hutinet[a], Walter Vetter[c], Giulia Poma[b], Yukiko Fujii[b,d], Robin E. Dodson[e], Boris Johnson-Restrepo[f], Dudsadee Muenhor[g,h,i], Bruno Le Bizec[a], Gaud Dervilly[a], Adrian Covaci[b], Ronan Cariou[a]*

[a]Oniris, INRAE, LABERCA, 44307 Nantes, France

[b]Toxicological Centre, University of Antwerp, 2610 Wilrijk, Belgium

[c]University of Hohenheim, Institute of Food Chemistry, 70599, Stuttgart, Germany

[d]Daiichi University of Pharmacy, Fukuoka, Japan, 815-8511

[e]Silent Spring Institute, Newton, MA, USA, 02460

[f]Environmental Chemistry Research Group, School of Exact and Natural Sciences, Campus of San Pablo, University of Cartagena, Cartagena 130015, Colombia

[g]Faculty of Environmental Management, Prince of Songkla University, Hat Yai, Songkhla 90110, Thailand

[h]Health Impact Assessment Research Center, Prince of Songkla University, Hat Yai, Songkhla 90110, Thailand

[i]Center of Excellence on Hazardous Substance Management (HSM), Bangkok 10330, Thailand

*Corresponding author

# Abstract

Bromochloro alkanes (BCAs) have been manufactured for use as flame retardants for decades and preliminary environmental risk screening suggests they are likely to behave similarly to polychlorinated alkanes (PCAs), subclasses of which are restricted as Stockholm Convention Persistent Organic Pollutants (POPs). BCAs have rarely been studied in the environment, though some evidence suggests they may migrate from treated-consumer materials into indoor dust, resulting in human exposure via inadvertent ingestion. In this study, BCA-$C_{14}$ mixture standards were synthesized and used to validate an analytical method. This method relies on chloride-enhanced liquid chromatography-electrospray ionization-Orbitrap-high resolution mass spectrometry (LC-ESI-Orbitrap-HRMS) and a novel CP-Seeker integration software package for homologue detection and integration. Dust sample preparation via ultrasonic extraction, acidified silica clean-up and fractionation on neutral silica cartridges was found to be suitable for BCAs, with absolute recovery of individual homologues averaging 66 to 78% and coefficients of variation ≤10% in replicated spiking experiments (n=3). In addition, a total of 59 indoor dust samples from six countries including Australia (n=10), Belgium (n=10), Colombia (n=10), Japan (n=10), Thailand (n=10) and the United States of America (n=9) were analysed for BCAs. BCAs were detected in seven samples from the USA, with carbon chain lengths of $C_8$, $C_{10}$, $C_{12}$, $C_{14}$, $C_{16}$, $C_{18}$, $C_{24}$ to $C_{28}$, $C_{30}$ and $C_{31}$ observed overall, though not detected in samples from any other countries. Bromination of detected homologues in the indoor dust samples ranged from $Br_{1-4}$ as well as $Br_7$, while chlorine numbers ranged from $Cl_{2-11}$. BCA-$C_{18}$ were the most frequently detected, observed in each of the USA samples, while the most prevalent halogenation degrees were homologues of $Br_2$ and $Cl_{4-5}$. Broad estimations of BCA concentrations in the dust samples indicated that levels may approach those of other flame retardants in at least some instances. These findings suggest that development of quantification strategies and further investigation of environmental occurrence and health implications are needed.

# Introduction

Chlorinated paraffins (CPs), mostly composed of polychlorinated alkanes (PCAs), are a complex mixture of tens of thousands of individual compounds which have been used as plasticizers, flame retardants and lubricants in a wide variety of consumer goods and materials including polyvinyl chloride, rubber, adhesives, sealants and textiles.[1, 2] Global production of PCAs has increased substantially during recent decades with recent total manufacture predicted to exceed one million tonnes per year as of 2020.[1] PCAs may be released from materials via volatilization, abrasion or direct transfer to other matrices[3] and have been identified as one of the major plasticizer contaminants in indoor dust, resulting in human exposure via inadvertent dust ingestion.[3-5] Commercial CP products are typically categorised by the most prominent carbon chain length of PCA constituents as short- ($C_{10-13}$), medium- ($C_{14-17}$) and long- ($C_{\geq 18}$) chain CPs (SCCPs, MCCPs and LCCPs, respectively).[6] SCCP, MCCP and LCCP groups have each exhibited bioaccumulative[7] and toxic endocrine disrupting properties[8-10] and have been detected globally in human blood and breastmilk in numerous studies. SCCPs have been subjected to legislated manufacture and usage restrictions since the mid-2000s in the EU[11, 12] and a number of other countries,[13-15] and were registered to the United Nations Stockholm Convention on Persistent Organic Pollutants (POPs) in 2017.[16] Addition of MCCPs to the Stockholm Convention has also been officially proposed and is currently being reviewed.[17]

A recent comprehensive market analysis indicated that global production and usage of SCCPs and MCCPs began to decline around 2014, with particularly pronounced reductions of SCCPs occurring in Western Europe, North America and the World's leading producer, China (Chen et al. 2022). This raises the question as to which compounds are replacing SCCPs and MCCPs in consumer products previously treated with these substances. While much of the demand may, indeed, be met by continued usage of LCCPs, the longest chained PCAs do not share exactly the same functionality as the SCCPs and MCCPs. Another potential candidate is the application of replacement bromochloro alkanes (BCAs), essentially PCAs with the addition of at least one bromine atom. Chemical analysis has showed that a substance registered as "$C_{12-30}$ bromo-chloro alkenes" (CAS: 68527- 01-5) in a commercial mixture marketed as "Doverguard 8207A", in fact, consisted predominantly of $C_{18}$ alkanes substituted with 1-3 bromine atoms and 3-7 chlorines.[18] It was hypothesized by Chibwe, *et al.* [18] that "$C_{12-30}$ bromo-chloro alkenes" are so misnamed due to the linear alpha-alkenes which are used as starting reagents but which are most likely converted to alkanes by halogenation at the terminal carbon position. Substances registered as "$C_{12-30}$ bromo-chloro alkenes" have been classified as high production chemicals in the United States of America (USA) since the 1980s and had a reported production volume between 45-227 t/y during the period of 2012-2015.[19] $C_{12-30}$ bromo-chloro alkenes are also listed in the chemical inventories of Australia, Canada, China, Korea and the European Commission's Registration, Evaluation, Authorisation and Restriction of Chemicals (REACH) preregistration list, although no production or import data is available.[18, 20] Numerous examples of commercial BCA products have been marketed under tradenames such as Doverguard-8207-A, -8408, -8208-A, Paroil 63-NR and are currently for sale as "Alkanes, $C_{10-18}$, bromo chloro" in China. Although no toxicological data is available, estimated physicochemical properties of BCAs indicative of their biological and environmental behaviour such as octanol-water, octanol-air and air-water partition coefficients ($K_{OW}$, $K_{OA}$ and $K_{AW}$, respectively), fall within the same ranges as PCAs.[18, 21]

The tremendous challenges inherent in analysing PCAs are well documented,[22-25] and relate to the substantial number of possible isomers represented by the general PCA formula, $C_xH_{(2x+2-y)}Cl_y$ for carbon chains which have ranged $C_{6-48}$ in environmental matrices.[3, 26] Complications derive from the fact that the PCA isomers cannot be fully separated using liquid or gas chromatography, while mass spectral isotopic interferences between homologues are also numerous.[27] Similar difficulties can be expected during analysis of BCAs, for which the composition of $C_xH_{(2x+2-y-z)}Br_yCl_z$ will include additional mass spectral complexity arising from the combination of both chlorine and bromine isotopologues. While 1728 elemental compositions are possible for PCAs ranging $C_{10-25}$, Li et al.[21] showed that the same carbon chain length range gives rise to 32,280 potential compositions for BCAs. For both BCAs and PCAs each elemental composition may consist of hundreds or thousands of isomers. Among the most powerful tools for discriminating isotopic features of PCAs and BCAs are high-resolution mass spectrometers (HRMS), such as quadrupole-time-of-flight (QTOF) and Orbitrap HRMS, which also provide high mass accuracy. An MS resolution exceeding 50,000 has been prescribed for discrimination of PCA isotope patterns,[27, 28] while Chibwe, et al.[18] recommended a minimum resolution of 20,000 to minimize mass spectral interferences between BCA homologues. Given these challenges, another approach for improving the identification of BCA homologues is to consider the whole of the isotopologue pattern of diagnostic ions for comparing the relative distribution and mass in precisely measured spectra with exact theoretical values, rather than rely upon ratios between just two or three *m/z* values. This is especially important for accurate detection of BCAs while no commercial analytical standards are available for use as reference materials and quantification calibrants.

This research addresses the lack of available authentic standards via synthesis of BCA mixtures with defined carbon chain lengths and varied degrees of bromination and chlorination, which were applied for validation of extraction methods and characterization of BCA ionization behaviour during analysis by LC-Orbitrap-HRMS. This study also aimed to determine the occurrence of BCAs in indoor dust collected from six countries, including Australia, Belgium, Colombia, Japan, Thailand and the USA, to provide a broad account of BCA contamination in indoor environments across diverse global regions. This research introduces a novel, custom-built "CP-Seeker" software package for automated identification of PCAs and related chemical families including BCA homologues, based on the integration of all detected isotopomer groups for enhanced selectivity during data interpretation and evaluates its utility for suspect screening of BCAs in indoor dust samples.

## Materials and methods

### Synthesis of bromochloro alkane standards

Four unique BCA-$C_{14}$ mixture standards were synthesised with varying degrees of bromination and chlorination using methods described in detail by Vetter, et al. [29,] Vetter, et al. [30]. Two separate reaction mixtures, designated A and B, were prepared with different Br and Cl ratios. Mixture A contained 86.6 mg bromine (1.08 mmol), 2501.1 mg of sulfuryl chloride (37.06 mmol), 1.0 mL dichloromethane and 0.1 mL of *n*-tetradecane (0.38 mmol). Mixture B was prepared with the same amounts of all constituents except for bromine, which was increased to 121.4 mg (1.52 mmol). The mixtures were placed in a beakers, topped with glass and irradiated with medium-pressure mercury vapor lamp. Aliquots were taken from each mixture after reaction times of 110 min (herein referred to as A1 and B1) and 220 min (A2 and B2). The organic layer was removed carefully washed, dried overnight, filtrated and condensed.

### Dust sample collection and preparation

A total of 59 indoor dust samples were obtained from archived collections with the objective of screening indoor environments for BCAs from a broad range of countries. All samples were collected between the years 2016 and 2019 from homes in Melbourne, Australia (n=10), Flanders, Belgium (n=10), Medellín, Colombia (n=10), Fukuoka Prefecture, Japan (n=10), and Kalasin Province, Thailand (n=10). Dust samples were also obtained from classrooms, lounge areas and auditoriums of college campuses in New England, USA (n=9). Domestic vacuum cleaners were used to collect settled dust from each of the countries and all samples were sieved to a <500 μm fraction and stored in darkness prior to analysis. Full details of the sampling protocols are provided in Section S1 of the Supplementary Information.

Extraction of BCAs was performed according to methods previously validated for extraction of PCAs from indoor dust.[31] Aliquots of 50 mg of dust were weighed into 15 mL glass vials and spiked with 20 ng of β-1,2,5,6,9,10-hexabromo[$^{13}C_{12}$]cyclododecane ($^{13}C$-β-HBCDD) for use as internal standard (IS). Samples were vortexed for 1 min in 5 mL of *n*-hexane and dichloromethane (3:1, v/v) and extracted by ultrasonication for 10 min. Vials were centrifuged for 3 min at 2000 rpm, the supernatants transferred to clean vials and the extraction repeated once more with clean solvent. Purification of the extracts was performed by addition of 2 g of acidified silica (44% w/w $H_2SO_4$) to vials and vortexing for 1 min. A second clean-up step was employed to separate organohalogen contaminants which may interfere with the analysis, entailing fractionation on Agilent Bond Elut silica (500 mg) cartridges. After loading samples to the cartridges, the first elution using 6 mL *n*-hexane was discarded and the second 12 mL dichloromethane fraction retained for analysis. The dichloromethane fraction was then concentrated to near dryness under nitrogen stream and reconstituted in 100 µL of acetonitrile containing [$^2H_{18}$]-γ-1,2,5,6,9,10-hexabromocyclododecane ($^2$H-γ-HBCDD) recovery standard (RS) for LC analysis. Full details of the reagents and standards are provided in Section S2.

*Data acquisition*

LC-HRMS analysis was carried out on an Ultimate 3000 LC coupled to a Q Exactive Orbitrap (Thermo Fisher) using electrospray ionization (ESI). Briefly, 5 µL sample extract injections were introduced to a Hypersil Gold column (100 mm × 2.1 mm, 1.9 µm) (Thermo Fisher) maintained at 30 °C. The mobile phase consisting of water and acetonitrile was held at 70% acetonitrile for 1 min and then evolved to 100% acetonitrile from 1 to 7 min, followed by a 10 min hold at 100% acetonitrile. The mobile phase then returned from 100% to 70% acetonitrile during 1 min before equilibrating for 2 min at a consistent 70% acetonitrile. The flow rate was 0.4 mL/min. A post-column T-connection was used to supply a mixture of acetonitrile and dichloromethane (1:1 v/v) at a flow rate of 0.08 mL/min to enhance the formation of [M + Cl]$^-$ ions. The ESI probe was operated at a voltage of 2.5 kV and capillary temperature of 275 °C, while the sheath gas and auxiliary gas flow rates were 50 and 5 arbitrary units, respectively. The Orbitrap-HRMS was operated in full scan mode from *m/z* 300 to 1500 with a resolution of 140,000 FWHM at *m/z* 200 and automatic gain control (AGC) target of 1 × 10$^6$.

*CP-Seeker data treatment*

BCA suspect screening and integration of mass spectral features was accomplished using the upgraded custom-built CP-Seeker v2.1.0 software package. Automated mass spectral searches were conducted for BCA homologues with the chemical formula $C_xH_{(2x+2-y-z)}Br_yCl_z$ ranging from $C_{6-36}$, $Br_{1-30}$ and $Cl_{1-30}$ (3 ≤ y + z ≤ x + 3) according to theoretical mass and isotopic patterns for ions of the form [M + Cl]$^-$ and [M – H]$^-$. For selected samples, [M + Cl]$^-$ ions of bromochloro olefins (BCOs) with formula $C_xH_{(2x-y-z)}Br_yCl_z$ or $C_xH_{(2x-2-y-z)}Br_yCl_z$ ranging from $C_{6-36}$, $Br_{1-30}$ and $Cl_{1-30}$ (3 ≤ y + z ≤ x + 3) and PCAs with formula $C_xH_{(2x+2-z)}Cl_z$ ranging from $C_{6-36}$ and $Cl_{3-30}$ (z ≤ x + 3) were also sought and integrated. Feature integration parameters were set to include peak widths ranging 5 to 300 s with a maximum of 20 missing scans, while BCA, BCO and PCA homologues were considered to be detected when the measured isotopic pattern matched theoretical ratios with a score ≥ 80% and weighted mass deviation ≤ 2 mDa. Using CP-Seeker, both the isotopic pattern match score and mass deviation are calculated from all isotopomers with a relative abundance greater than 1% of the base-peak for the targeted ion, ensuring enhance identification capacity. The CP-Seeker software generated a single response value per BCA homologue comprising the sum of detected isotopomer group peak areas for each selected ion to account for signal dispersion due to isotope combinations. A broader description of the CP-Seeker function and operation is provided in Section S3, Figures S1–10 and Table S1. The application is freely available upon request at contact.cpseeker@oniris-nantes.fr, under the CC-BY 4.0 license, and is delivered with a comprehensive user guide documentation.

## *Quantification of PCAs*

Quantification of $\Sigma PCA\text{-}C_{10\text{-}13}$, $\Sigma PCA\text{-}C_{14\text{-}17}$ and $\Sigma PCA\text{-}C_{18\text{-}20}$ was performed for selected samples using the pattern reconstruction procedure as described by McGrath, et al.[32], based on previously published methods.[33,34] Briefly, six-point calibration curves were constructed from technical mixture standards of $\Sigma PCA\text{-}C_{10\text{-}13}$ (51.5, 55.5 and 63 %Cl), $\Sigma PCA\text{-}C_{14\text{-}17}$ (42, 42 and 57 %Cl) and $\Sigma PCA\text{-}C_{18\text{-}20}$ (36 and 49 %Cl) and relative mean responses calculated for PCAs by dividing PCA homologue mean responses by the response of the $^{13}C\text{-}\beta\text{-}HBCDD$ internal standard. PCA homologue profiles in the samples were reconstructed from linear combinations of the patterns in standards to determine appropriate mean response factors for $\Sigma PCA\text{-}C_{10\text{-}13}$, $\Sigma PCA\text{-}C_{14\text{-}17}$ and $\Sigma PCA\text{-}C_{18\text{-}20}$ quantification. Quantification of BCAs and BCOs was not performed as appropriate, well-characterized standards were not available for these compound classes.

## *Quality assurance and quality control*

Spike and recovery tests were conducted to confirm that the application of extraction, clean-up and analysis protocols previously validated for PCAs were also suitable for BCAs. Synthesized $BCA\text{-}C_{14}$ standards A2 and B2 were selected for recovery experiments as they represented the mixtures with the lowest and highest bromination degree, respectively. A2 and B2 standards were spiked into separate empty 15 mL glass vials, in triplicate, at a rate of 100 µL of 1 µg/mL $\Sigma BCA\text{-}C_{14}$ (representing a concentration of 2 µg/g in dust) and processed according to the full sample preparation and analysis procedure (final volume 100 µL). Individual homologue recoveries were calculated as the mean response in spiked sample analyses (n=3), divided by response in the spike mixture used for fortification (1 µg/mL). Ongoing extraction efficiency was monitored by the recovery of the $^{13}$C-β-HBCDD IS as corrected against response of the $^{2}$H-γ-HBCDD RS, which averaged 82 % with a range of 55 to 126 % across all dust and QA/QC samples. The IS was also used to derive relative responses by dividing the response of BCA homologues by the respective responses of IS to correct for analytical variation between injections. A procedural blank was prepared with each batch of 15 sample extractions (total n=4) and two field sampling blanks were analysed from each of the countries except for Thailand, where field blanks were not available. Details of the field blank collection are provided in Section S1. BCAs were not detected in any blanks. Only PCA-C$_{14}$ were detected in field or procedural blanks, with a mean total response which was exceeded by the lowest PCA-C$_{14-17}$ calibration point by a confidence interval of 95%.

## Results and discussion

*Characterization of BCA-C$_{14}$ standards*

The synthesised mixture standards were characterized by LC-HRMS using injections of BCA-C$_{14}$ at concentrations of 10 µg/mL. BCA-C$_{14}$ eluted between 3.6 and 6.5 min as broad peaks similar to those of PCAs analysed under the same conditions (Figure 1B-D). The post-column addition of dichloromethane is often applied in PCA analysis by LC-ESI or LC-APCI-HRMS to favour [M + Cl]$^-$ formation over [M – H]$^-$ to improve sensitivity and reduce potential mass spectral interferences.[27] This approach was applied for BCA analysis in the current study and [M + Cl]$^-$ ions were detected for BCA-C$_{14}$ homologues with Br$_{1-7}$ and Cl$_{2-8}$ and PCA-C$_{14}$ (i.e. Br$_0$) with Cl$_{4-8}$ among the four standards. [M – H]$^-$ ions were not detected in any of the mixtures for BCAs or PCAs. Figure 2 shows the relative mean responses measured for individual homologue groups of both BCA-C$_{14}$ and PCA-C$_{14}$ in the standards, normalised to the maximum mean response per standard. By this measure, bromine patterns were similar for the A1 and A2 standards with relative responses of Br$_{1-2}$ accounting for a combined 54 and 55 %, respectively, with Br$_3$ homologues contributing 18 and 14 %, respectively, and Br$_4$ and Br$_5$ each ≤ 8 %. The B1 standard was dominated by each of the Br$_1$, Br$_2$ and Br$_3$ homologues (19, 26 and 25 %, respectively) followed by Br$_4$ (14 %) and Br$_5$ (6 %), while the B2 standard was dominated by Br$_2$, Br$_3$ and Br$_4$ (24, 28 and 21 %, respectively) followed by Br$_1$ (14 %) and then Br$_5$ (6 %), Br$_6$ (2 %) and Br$_7$ (1%). The most abundant Cl groups in the A1, A2, B2 and B2 mixtures were Cl$_{4-5}$, Cl$_{5-6}$, Cl$_{3-4}$ and Cl$_{4-5}$, respectively. Variation in the BCA chromatographic peak shapes and retention times between the standard mixtures was observed for peaks of the same elemental composition, suggesting that the isomeric composition of individual homologues also differed between the standards (Figure S10). PCAs detected in the standard mixtures were also prominent, with higher proportions of PCAs generated from reaction mixture A, featuring the lower bromine ratio. ΣPCA-$C_{14}$ relative responses equating to approximately 19, 26, 10 and 5 % of the total combined response of ΣPCA-$C_{14}$ and ΣBCA-$C_{14}$ were observed in the A1, A2, B1 and B2 standards, respectively. Mass fractions of bromine based on the homologue groups detected by LC-HRMS measurement were 25.1, 20.4, 34.1 and 36.2 % w/w in the A1, A2, B1 and B2 standards, respectively, reflecting the higher ratio of bromine employed in the synthesis of the B mixtures. Chlorine mass fractions in the standards were also affected by the Br/Cl synthesis ratios, with higher chlorination degrees observed in A1 and A2 (35.2 and 41.3 %w/w, respectively) than B1 and B2 (26.9 and 29.4 % w/w, respectively). Overall LC-HRMS response factors were highest in the standards produced from the longer reaction time, with the relative response of ΣBCA-$C_{14}$ greater for each of the A2 and B2 mixtures than those of A1 and B1 (Figure 1A). ΣPCA-$C_{14}$ relative responses were also greatest in the mixtures produced with lower ratios of bromine, with A2 showing the greatest response followed by A1, B2 and B1.

The occurrence of chemical constituents other than BCA-$C_{14}$ and PCA-$C_{14}$ were also investigated in the mixture standards. The acquired full-scan LC-ESI-HRMS chromatograms were extensively searched for halogenated (Br and Cl) ions using HaloSeeker v2.0.3.3, a software developed to specifically screen halogenated organic molecules on the basis of chlorine and bromine isotopic ratio and mass defect.[35, 36] The complete procedure and results are provided in Section S4, Figure S11 and Table S2. While BCAs and PCAs of carbon chain lengths other than $C_{14}$ were not detected in the mixtures, series of more polar mixed polyhalogenated compounds composed of isomer mixtures were identified, which may have arisen from uncontrolled sulphuryl chloride adduct formation during synthesis, as has been reported previously.[37] Although neither exhaustive nor quantitative, this characterization provided some rough information on the purity of the BCA-$C_{14}$ mixture standards, which appeared to be much higher for the mixture standards A2 and B2.

Spike and recovery tests performed using the A2 and B2 standards showed the extraction and clean-up protocol to be appropriate for sample preparation with absolute recoveries of individual BCA-$C_{14}$ homologues averaging 66 to 78 %, each with coefficients of variation (CV) ≤ 10 %. Recoveries of individual BCA-$C_{14}$ homologues are presented in Table S3. Recovery of the $^{13}$C-β-HBCDD IS were similar to BCAs with mean ± CV recoveries of 71 ± 7 % and 77 ± 7 % for the A2 and B2 tests, respectively. These findings indicated that the $^{13}$C-β-HBCDD was a suitable IS for deriving BCA relative responses measured in dust samples to account for extraction losses, as well as analytical variation during comparisons between injections.

*Occurrence of BCAs and BCOs in indoor dust*

BCAs were detected in a total of seven indoor dust samples among the six countries, all of which were from the USA (Tables S4–6). Figure S12 provides an example mass spectrum from dust sample US-6 showing discrimination of the [M + Cl]$^-$ M+2 peak of BCA-$C_{12}H_{21}Br_2Cl_3$ from other compounds at a resolution of ~97,000. This is consistent with predictions by Li, *et al.* [21] that an MS resolution of 60,000 would be sufficient to distinguish most of the likely mixed halogenated contaminants from PCAs when coupled with chromatography. Overall, BCAs of carbon chain lengths $C_8$, $C_{10}$, $C_{12}$, $C_{14}$, $C_{16}$, $C_{18}$, $C_{24}$, $C_{25}$, $C_{26}$, $C_{27}$, $C_{28}$, $C_{30}$ and $C_{31}$ were detected in indoor dust, with distribution patterns differing greatly between samples. BCA-$C_{14}$ homologues were only detected in one of the dust samples (US-6) to allow for direct comparison with synthesised standards, showing chromatographic variation indicative of distinct isomeric compositions between this sample and the standards (Figure S10). A visual comparison of measured and theoretical isotopic patterns of selected BCA homologues for sample US-6 is also provided in Figure S10. BCA-$C_{18}$ were the most frequently detected, observed in each of the seven USA samples, while $C_{12}$, $C_{26}$ and $C_{30}$ BCAs were observed in three samples each, $C_8$, $C_{10}$, $C_{24}$ and $C_{28}$ BCAs in two samples each and the remaining carbon chain length groups in only single samples. With respect to relative response, the most prominent of the chain length groups was $C_{12}$ followed by $C_{14}$ and then $C_{18}$ (Figure 3). Among the samples in which BCAs were detected, relative responses for carbon chain lengths other than $C_{12}$, $C_{10}$ and $C_{18}$ homologues were proportionally much lower. A prominence of even carbon-chained BCAs was also apparent, which may relate to distinct alkane feedstocks utilized in the manufacture of BCA flame-retardant formulations, and has also been observed for PCAs in indoor dust previously.[4] Bromination of detected homologues in the indoor dust samples ranged from $Br_{1-4}$ as well as $Br_7$, while chlorine numbers ranging from $Cl_{2-11}$ were observed. Homologue groups containing $Br_2$ were detected in each of the seven USA samples, $Br_1$ BCAs were detected in five samples, $Br_3$ and $Br_7$ homologues were each observed in three samples and only one $Br_4$ homologue group was detected in a single sample. $Cl_4$ and $Cl_5$ were most prominent in the BCA groups among the dust samples followed by $Cl_6$ and $Cl_7$ homologue groups. BCOs were investigated for the samples in which BCAs were identified and were only detected in sample US-6, as eight monounsaturated $C_{10}$ and $C_{12}$ homologue groups containing $Br_{1-2}$ and $Cl_{3-7}$ (Table S7). The relative abundance of the $C_{10}$ and $C_{12}$ homologue groups was 12 and 88%, while the combined relative response of all detected BCOs accounted for only 2.3% of the relative response of total detected BCAs in sample US-6. BCOs may represent impurities or transformation products formed during manufacture of the primary or secondary material products.

To the authors' knowledge, analysis of BCAs and BCOs in environmental samples has only been reported in one previous study. He, et al. [38] detected both BCAs and BCOs in ≥ 90 % of indoor dust samples from 44 homes, 10 offices and seven public transport vehicles in Australia via LC-atmospheric pressure chemical ionization (APCI)-QTOF analysis, in contrast with the present study having detected neither compound class in Australian dust samples. This discrepancy may arise from differences in the consumer goods or construction materials of individual sample locations, as well as the fact that samples were collected from separate cities in different states (Melbourne, Victoria in the present study, versus Canberra in the Australian Capital Territory and Brisbane, Queensland). Differences in instrumental sensitivity and detection criteria between the studies may also partly account for the discrepancy, since estimates of limits of detection for BCAs and BCOs cannot be reliably derived, while well-characterised standards are not available. BCA homologue profiles reported by He, et al. [38] were broadly similar to those detected in the USA samples of this study, with carbon chains ranging $C_{10-21}$, $Br_{1-6}$ and $Cl_{1-10}$, although the prominence of even carbon-chain lengths among samples of the current study were not observed.

The greatest number of individual BCA homologue groups, 37, were detected in sample US-6, followed by US-4 (29), US-5 (15), US-2 (12), US-1 (10), US-3 (3) and US-7 (2). While for most samples, too few BCA homologue groups were detected to observe clear patterns among relative mean responses, approximately Gaussian distributions of the number of bromine and chlorine atoms in BCAs were discerned for sample US-6 (Figure 4), reflecting the general patterns detected in the synthesised standards. Such Gaussian distributions in bromination and chlorination were also observed in LC-APCI-Orbitrap-MS analyses of a commercial bromochloro alkane formula with the tradename Doverguard 8207A, in which $Br_{1-3}$ and $Cl_{3-7}$ were dominant.[18]

Homologue patterns distinct from those observed for BCAs in the range of $C_{8-18}$ in the dust samples and synthesised standards were apparent in samples US-1 and US-5. In these samples, a series of BCAs with chain-lengths ranging $C_{24-28}$ and chlorination of $Cl_{5-11}$ were each detected as homologues of only $Br_7$ (Tables S4–6, Figure S13). BCA homologues of $Br_6$ or $Br_8$, whose occurrence may be predicted based on the Gaussian bromine substitution profiles observed in the synthesized standards and Doverguard 8207A formula,[18] were not detected for any of the $C_{24-28}$, $Cl_{5-11}$ groups. Further manual inspection of the acquisition data revealed mass spectral features relating to the expected $Br_6$ and $Br_8$ BCA patterns with mass deviation < 3 mDa for both US-1 and US-5, which were not automatically integrated by the CP-Seeker software. The retention times and relative responses of these features were consistent with the presence of bromine series dominated by $Br_7$ and smaller proportions of $Br_6$ and $Br_8$ at the observed BCA-$C_{24-28}$ chain lengths for both samples. It is likely that these mass spectral features were not detected by CP-Seeker as they failed to meet the peak identification criteria with the embedded xcms package. Samples US-1 and US-5 were not collected from the same campus and further investigation would be required to determine whether similar materials were acting as point-sources of these specific BCA profiles in each location.

Given the low detection rate for BCAs among the sample set as a whole, it is difficult to discern potential sources to dust or identify contributing factors for BCA contamination. The detection frequency of 7 out of 9 samples (~77 %) from the USA may suggest that manufacture and/or application of BCAs is more prominent in the USA than in the other studied regions. The broader detection frequency of BCA-$C_{18}$ in the USA indoor dust samples corresponds with BCA-$C_{18}$ having been identified as the major constituent of the Doverguard 8207A technical formula from Dover Chemical Corporation in Ohio, USA.[18] The classification of BCAs as high production chemicals in the USA (45-227 t/y during 2012-2015) [19] also supports these findings, although BCA manufacture or import volumes in the other countries of this study are not currently available. Another distinguishing feature of the US dust samples is that they were collected in public spaces (college buildings), while all other samples were obtained from private residences. The rooms from which dust containing BCAs were collected included auditoriums, lecture halls and lounge areas which were carpeted (except for US-5) and generally densely fitted with upholstered furnishings, ranging from 18 furnishings in US-5 to 462 in US-4. Previous studies have reported that contamination levels of brominated flame retardants (BFRs) in indoor dust from campuses in New England, USA, correlated with legislative changes in furniture flammability standards.[39,40] Specific laws in the USA may have influenced the flame-retardant constituents utilized in the interior furnishings of educational institutions, in turn affecting BCA concentration in dust.

*Comparison between BCAs and PCAs in indoor dust*

PCA concentrations in the dust samples from Australia, Colombia, Japan and Thailand have been published previously[32] and levels in the Belgian samples were reported by McGrath*, et al.* [4]. Although no BCAs were detected in the vast majority of these samples, PCAs were identified in all samples from these countries at overall $\Sigma$PCA-$C_{10-13}$, $\Sigma$PCA-$C_{14-17}$ and $\Sigma$PCA-$C_{18-20}$ concentrations ranging 1.2 to 290 µg/g, 6.9 to 540 µg/g and <1.0 to 230 µg/g, respectively (Table S8). For comparative purposes, PCAs of $C_{10-13}$, $C_{14-17}$ and $C_{18-20}$ were also quantified in each of the USA samples in which BCAs were detected. Pattern reconstruction goodness of fit ($R^2$) was > 0.65 for all US sample measurements except for one ($R^2$ = 0.48 for $\Sigma$PCA-$C_{18-20}$ in sample US-5) (Table S9). PCA homologues ranging $C_{8-35}$ were detected in the US dust samples (Figure S14), with $\Sigma$PCA-$C_{10-13}$, $\Sigma$PCA-$C_{14-17}$ and $\Sigma$PCA-$C_{18-20}$ concentrations ranging from 14 to 53, 15 to 210 and 2.8 to 43 µg/g, respectively (Table 1). Comparison of BCA relative mean responses with those of PCAs quantified in this study may provide an approximation of the magnitude of BCA contamination, while well-characterized commercial standards of varied carbon chain length are unavailable. CP-Seeker inherently accounts for the fractional isotopic abundance differences between individual BCA and PCA homologues by deriving response output as the sum of areas integrated for all detected isotopomers of the target ion. Earlier research has shown a correlation between PCA homologue mean response and molecular mass with peak ionization efficiency in the range of around *m/z* 400 to 600 using LC-APCI-QTOF.[38, 41] Accordingly, comparison between BCAs and PCAs are presented by carbon chain length groups in Figure 5, and assume ionization efficiency to be approximately similar (within an order of magnitude).

The relative mean response of $\Sigma$BCA-$C_{6-9}$ was approximately a third that of $\Sigma$PCA-$C_{6-9}$ in sample US-4, but more than an order of magnitude greater in US-6. PCA-$C_{6-9}$ have been reported in indoor dust and air by a small number of studies[3, 42] and typically represent in very small proportions of total PCAs in technical formulas.[43] While PCA-$C_{6-9}$ measured in CP commercial products have generally been considered to be unintended impurities within commercial CP products,[43] it is not known whether BCA-$C_{6-9}$ are intentional constituents or by-products of manufacture. For samples in which BCA-$C_{10-13}$ were detected, the $\Sigma$BCA-$C_{10-13}$ relative mean response was approximately two orders of magnitude lower than $\Sigma$PCA-$C_{10-13}$ in sample US-3, but similar in US-4 and around 3 times higher in sample US-6. This may suggest that concentrations of $\Sigma$BCA-$C_{10-13}$ in US-4 and US-6 are within the low- to mid-µg/g range, based on the $\Sigma$PCA-$C_{10-13}$ concentrations of 51 and 53 µg/g, respectively, in these samples. $\Sigma$BCA-$C_{14-17}$ relative responses recorded in samples US-4 and US-6 were both around two orders of magnitude below those of $\Sigma$PCA-$C_{14-17}$ (72 and 32 µg/g, respectively) to suggest that the $\Sigma$BCA-$C_{14-17}$ concentrations in these samples may be in the mid- to high ng/g range. For $\Sigma$BCA-$C_{18-20}$, which comprised only $C_{18}$ homologue groups, relative responses were one to two orders of magnitude lower than those of $\Sigma$PCA-$C_{18-20}$ for US-1, US-3, US-5, US-6 and US-7, but only approximately three-fold lower for US-2 and US-4. With $\Sigma$PCA-$C_{18-20}$ concentrations of 43 and 10 µg/g recorded in the US-2 and US-4 samples respectively, a broad estimate could place $\Sigma$BCA-$C_{18-20}$ levels in the high ng/g or low µg/g ranges. Although only very general estimates of BCA concentration ranges can be made on the basis of PCA concentrations in samples, these values would suggest that BCA levels may approach those of other organic flame retardants, such as BFRs or organophosphate flame retardants (OPFRs)[39, 44, 45] in at least some indoor environments. It appears that most of the BCAs detected in this study are likely present at low concentrations with respect to PCA contamination levels.

## Conclusions

The results of this study indicate that BCAs are not likely to occur widely in indoor dust from homes in most of the studied countries but may be prevalent within indoor environments of the USA. The international screening approach of this research necessitated that relatively few samples could be analysed from separate locations, and a broader investigation of indoor dust from the USA is warranted to elucidate the contamination status of BCAs in public buildings and other indoor settings. The task of assessing BCA levels in dust remains very challenging, while analytical standards are unavailable. Although the estimated concentration ranges presented for BCAs in this study are subject to a high degree of uncertainty, it is reasonable to expect that the ΣBCA levels observed in USA samples constitute a considerable contribution to overall chemical exposures with respect to the levels often reported for other flame retardants such as BFRs, OPFRs and PCAs, in indoor dust. The high level of halogenation-specific variation between relative responses observed for the four BCA-$C_{14}$ standards synthesized in this study indicates that accurate quantification will be strongly reliant on profiles in standards matching closely with those of patterns in samples. Efforts directed toward the development of BCA analytical standards and quantification strategies are required for further assessment of BCA environmental occurrence and the investigation of potential health consequences of BCA exposure.

## Supporting Information

Supporting Information Available:

Detailed description of indoor dust sample collection and locations, details of chemicals and reagents, in-depth explanation of the CP-Seeker software functions and operation, example extracted ion chromatograms of BCAs in indoor dust, details of impurities characterization in BCA standards with mass defect plot, recoveries of individual BCA homologue groups during fortification experiments, mass spectra of example BCA homologue measured in indoor dust sample (US-6), Tables of normalised relative response, isotopic match score and mass deviation for all BCA homologue groups detected in indoor dust, results of BCOs detected in indoor dust, concentrations of PCAs measured in indoor dust sample, pattern reconstruction goodness of fit for PCA concentrations in US dust samples, normalised relative response figures of BCA homologues in dust samples US-1 and US-5, relative abundance of PCAs detected in US dust samples.

## Acknowledgements

Financial support was provided from postdoctoral fellowships for Thomas J. McGrath from the European Union's Horizon 2020 research and innovation programme under the Marie Skłodowska-Curie Actions project number 101110252 (ANEX-PXA) and Scientific Research Foundation-Flanders (Fonds Wetenschappelijk Onderzoek, FWO) – project number 12Z9320N. Further support provided by the Exposome Centre of Excellence of the University of Antwerp (BOF grant, Antigoon database number 41222). Dudsadee Muenhor is grateful to the Thailand-United States Educational Foundation (TUSEF/Fulbright Thailand) for the Fulbright Thai Visiting Scholar Program (TVS). B. Johnson-Restrepo would like to thank the Ministry of Science, Technology, and Innovation (MinCiencias) for support of the samples collected in Colombia under Grant No. 110759634967. Yukiko Fujii would like to thank the Japan Society for the Promotion of Science (JSPS) (Grant No. 21K12262). Thank you to Sina Schweizer, Clara Hägele and Tobias Schulz from the University of Hohenheim for their contributions in the synthesis of mixture standards. The authors also extend their gratitude to all participants for allowing the collection of dust samples in their homes.

# Tables and Figures

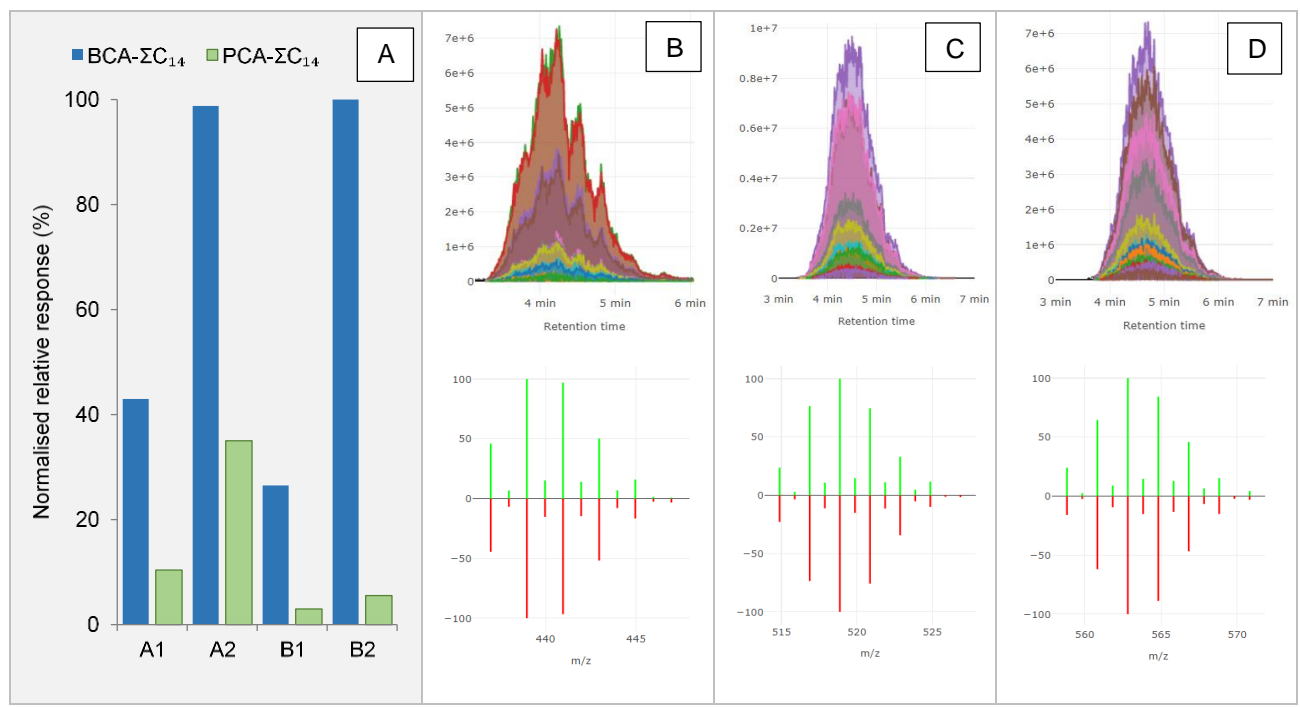

**Figure 1.** A: Normalised relative response (%) of $\Sigma$BCA-$C_{14}$ and $\Sigma$PCA-$C_{14}$ detected as $[M + Cl]^-$ in synthesized standards. B, C and D: Example chromatograms and mass spectra of $[M + Cl]^-$ for $C_{14}H_{24}Cl_6$, $C_{14}H_{23}BrCl_6$ and $C_{14}H_{23}Br_2Cl_5$ isotopomers, respectively, in synthesised standard A2, generated by CP-Seeker software. Chromatograms display overlaid results for each of the detected isotopomers. Mass spectra show measured profile above (green) and theoretical profile below (red). Mass spectra in Figures B, C, and D had isotopic pattern match scores of 98, 97 and 95%, respectively, and mass deviation of 0.58, 0.55 and 0.50 mDa, respectively, compared with theoretical values.

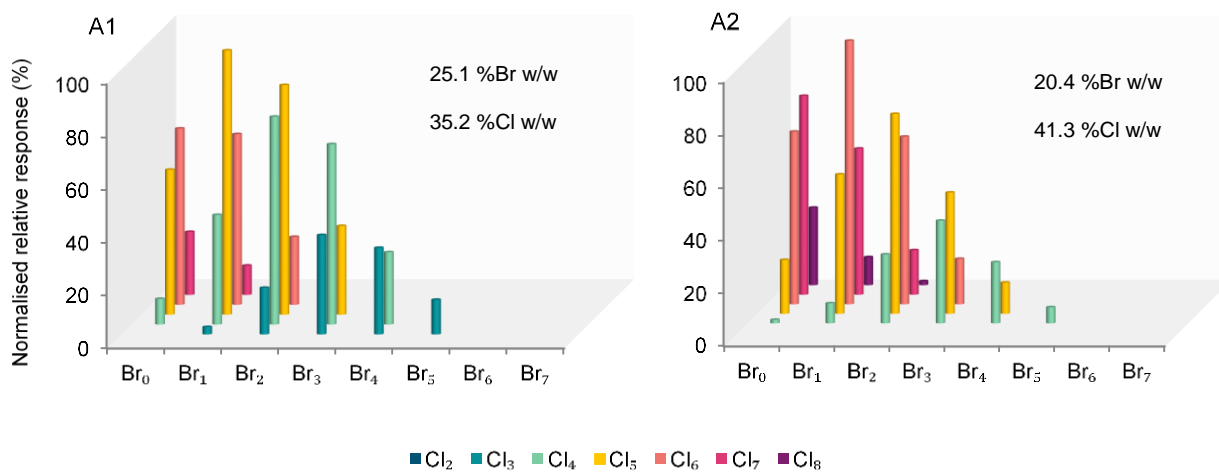

**Figure 2.** Normalised relative response (%) of bromine and chlorine on $C_{14}$ *n*-alkane chains of synthesised standards A1, A2, B1 and B2, normalised per individual standard. Br and Cl mass fractions (% w/w) indicated on Figures account for BCA and PCA ($Br_0$) homologue groups detected by LC-ESI-Orbitrap analysis.

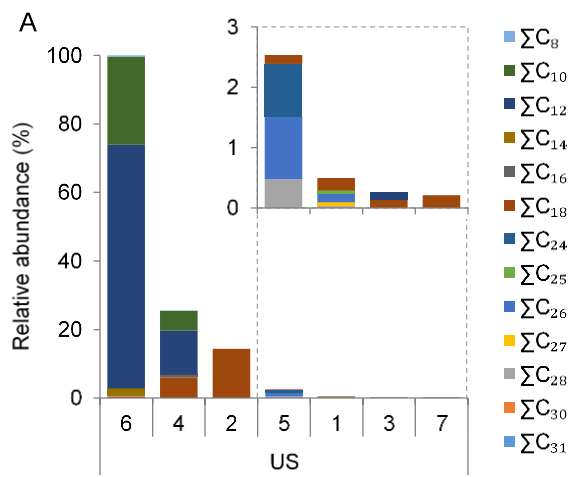

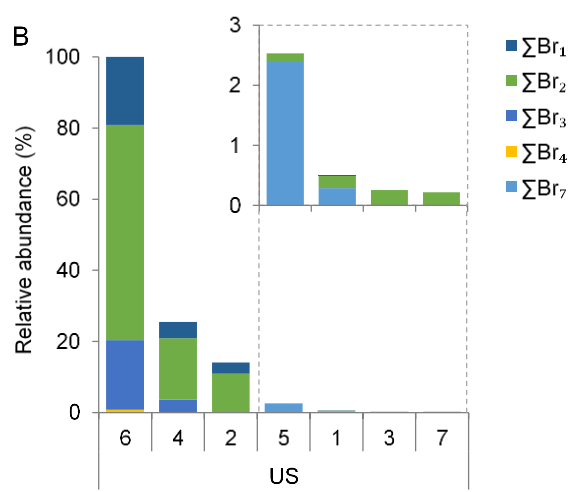

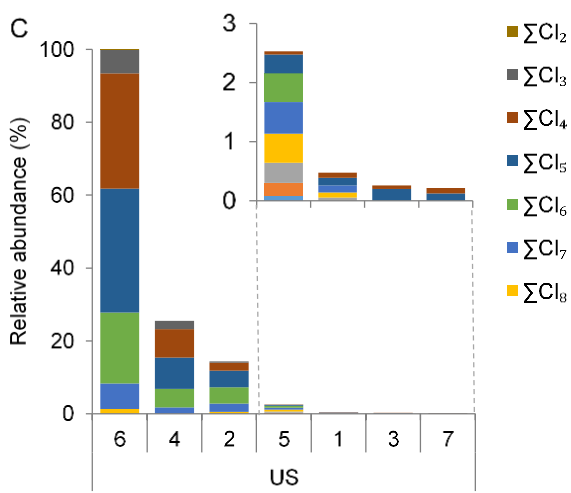

**Figure 3.** Normalised relative response (%) of BCA homologue groups detected in indoor dust samples grouped by A) carbon, B) bromine and C) chlorine numbers.

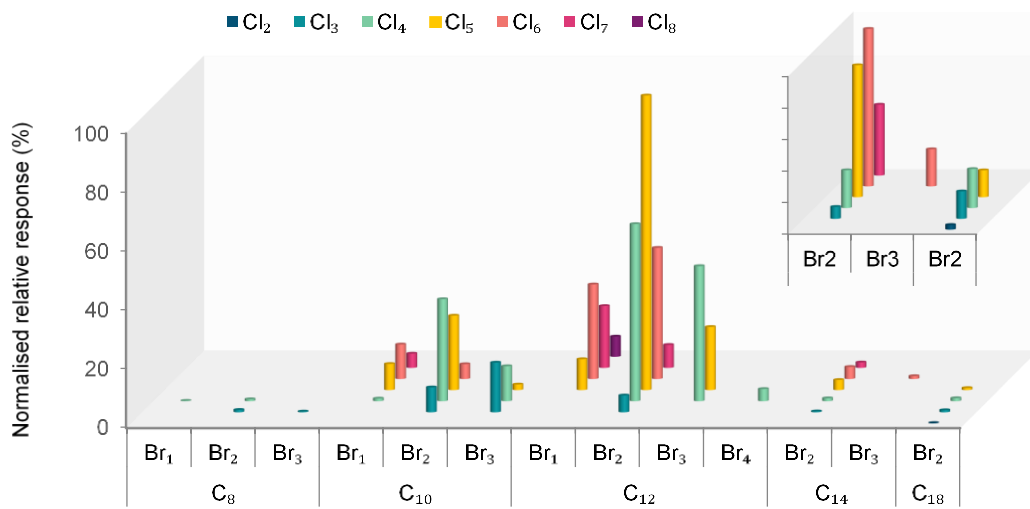

**Figure 4.** Normalised relative response of BCA homologue groups detected in indoor dust sample US-6.

**Table 1.** PCA concentrations (μg/g) in indoor dust samples US 1-7.

| Sample | ΣPCA-$C_{10-13}$ | ΣPCA-$C_{14-17}$ | ΣPCA-$C_{18-20}$ |
|---|---|---|---|
| US-1 | 14 | 210 | 10 |
| US-2 | 15 | 57 | 43 |
| US-3 | 34 | 52 | 7.5 |
| US-4 | 51 | 72 | 10 |
| US-5 | 4.8 | 15 | 2.8 |
| US-6 | 53 | 32 | 3.8 |
| US-7 | 24 | 53 | 9.4 |

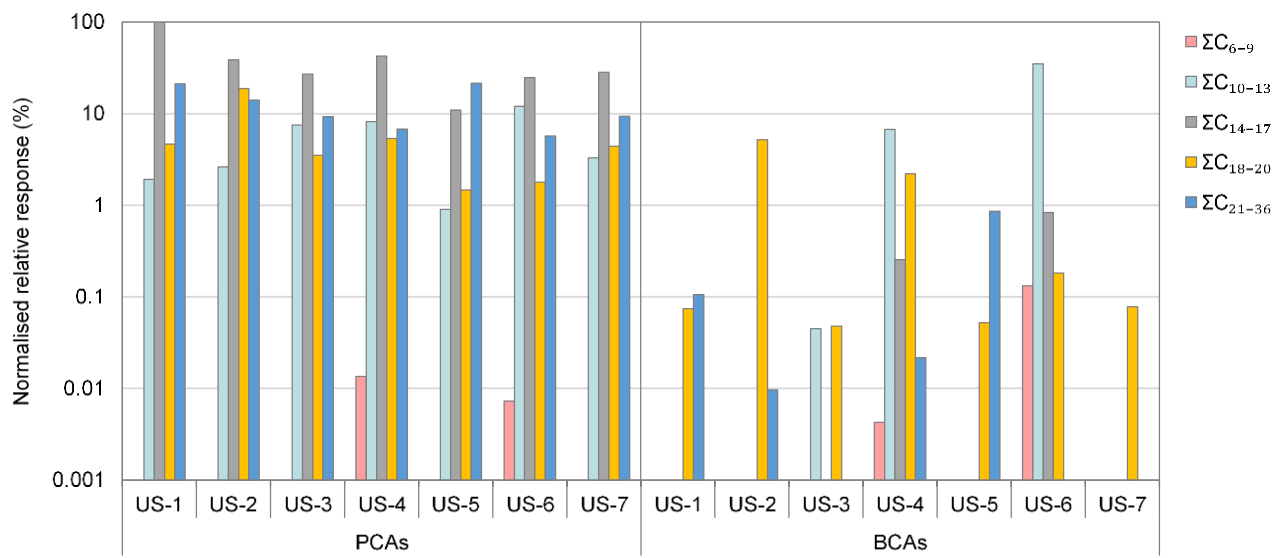

**Figure 5.** Normalised relative response (%) of PCAs and BCAs grouped by carbon chain length.